\documentclass[fleqn,usenatbib]{mnras}

\usepackage{newtxtext,newtxmath}
\usepackage[T1]{fontenc}

\DeclareRobustCommand{\VAN}[3]{#2}
\let\VANthebibliography\thebibliography
\def\thebibliography{\DeclareRobustCommand{\VAN}[3]{##3}\VANthebibliography}

\usepackage{graphicx}	% Including figure files
\usepackage{amsmath}	% Advanced maths commands
\usepackage{comment}
\usepackage{tablefootnote}
\usepackage{orcidlink}
\title[21cm PS with Pop. III star formation]{Semi-analytic modelling of Pop. III star formation and metallicity evolution - II. Impact on 21cm power spectrum.}

\author[E. M. Ventura et al.]{
Emanuele M. Ventura$^{\orcidlink{0000-0003-3502-4929}}$$^{1,4}$\thanks{E-mail: eventura@student.unimelb.edu.au},
Yuxiang Qin$^{\orcidlink{0000-0002-4314-1810}}$$^{2,4}$,
Sreedhar Balu$^{\orcidlink{0000-0002-5281-5151}}$$^{1,3,4}$,
and J. Stuart B. Wyithe$^{\orcidlink{0000-0001-7956-9758}}$$^{2,4}$
\\
$^{1}$School of Physics, University of Melbourne, Parkville, Victoria, Australia\\
$^{2}$Research School of Astronomy and Astrophysics, Australian National University, Canberra, ACT 2611, Australia\\
$^{3}$Facultad de Físicas, Multidisciplinary Unit for Energy Science, Universidad de Sevilla, 41012, Seville, Spain\\
$^{4}$ARC Centre of Excellence for All Sky Astrophysics in 3 Dimensions (ASTRO 3D)\\
}

\date{Accepted XXX. Received YYY; in original form ZZZ}

\pubyear{2024}

\begin{document}
\label{firstpage}
\pagerange{\pageref{firstpage}--\pageref{lastpage}}
\maketitle

\begin{abstract}
Simulating Population (Pop.) III star formation in mini-halos in a large cosmological simulation is an extremely challenging task but it is crucial to estimate its impact on the 21cm power spectrum. In this work, we develop a framework within the semi-analytical code \textsc{meraxes} to estimate the radiative backgrounds from Pop. III stars needed for the computation of the 21cm signal. We computed the 21cm global signal and power spectrum for different Pop. III models varying star formation efficiency, initial mass function (IMF) and specific X-ray luminosity per unit of star formation (L$_{\rm X}$/SFR). In all the models considered, we find Pop. III stars have little to no impact on the reionization history but significantly affect the thermal state of the intergalactic medium (IGM) due to the strong injection of X-ray photons from their remnants that heat the neutral IGM at $z \geq$ 15. This is reflected not only on the 21cm sky-averaged global signal during the Cosmic Dawn but also on the 21cm power spectrum at $z \leq$ 10 where models with strong Pop. III X-ray emission have larger power than models with no or mild Pop. III X-ray emission. We estimate observational uncertainties on the power spectrum using \textsc{21cmsense} and find that models where Pop. III stars have a stronger X-ray emission than Pop. II are distinguishable from models with no or mild Pop. III X-ray emission with 1000 hours observations of the upcoming SKA1-low.
\end{abstract} %249 words

\begin{keywords}
galaxies: high-redshift -- stars: Population III -- dark ages, reionization, first stars.
\end{keywords}

%%%%%%%%%%%%%%%%%%%%%%%%%%%%%%%%%%%%%%%%%%%%%%%%%%

%%%%%%%%%%%%%%%%% BODY OF PAPER %%%%%%%%%%%%%%%%%%

\section{Introduction}

When and where did Population III (Pop. III) stars form? What role did they play in the Cosmic Dawn and Epoch of Reionization (EoR)? And what is the best way to detect them? These questions remain open as no definitive observation of a mini-halo or Pop. III star has been reported.
%and only few candidates \citep[e.g.][]{Wang2024} have been reported. Up to the present day, we are relying on several cosmological simulations that investigate gas cooling and star formation in metal-free mini-halos ($M \sim 10^{5-7}$M$_\odot$).\\ 
To gain insight, small size and high-resolution hydrodynamical simulations have been performed \citep[e.g.][]{Greif2011, Hirano2018, Chon2021, Chon2022, Toyouchi2023, Sadanari2024} that suggest that metal-free (or poor) mini-halos favour the formation of Pop. III stars with a more top-heavy Initial Mass Function (IMF) and with lower star formation efficiencies than observed today. It is also thought that Pop. III star formation might occur down to the end of the EoR at $z \lesssim 6$ in pristine metal free pockets of gas \citep[e.g.][]{Venditti2023}.\\ 
The variety and complexity of the processes involved in Pop. III star formation and the resolution required to keep track of the evolution of the gas particles, limits the size of these hydrodynamical simulations to $\sim 100$ kpc. In order to mitigate this problem, semi-analytical models that account for Pop. III star formation have been developed \citep[e.g.][]{Visbal2020,Hegde2023,Boyuan2024}. These models allow a statistical study of Pop. III star formation in mini-halos out to scales of $\sim 10$ Mpc. While these volumes start to investigate the chemical enrichment of the intergalactic medium (IGM) and the Pop. III/II transition, they are still too small to study the EoR as volumes of at least $\sim 200$ Mpc are required (\citealt{Iliev2014, Kaur2020, Balu2023}).

Observations and models are converging on a scenario where the Universe was completely ionized by $z \sim 5.3$ \citep[e.g.][]{Fan2006, Ouchi2010, McGreer2015, Qin2021, Bosman2022,Qin2024} with reionization likely driven by low-mass halos \citep[e.g.][]{Kuhlen2012, Qin2021b, Saxena2024, Mutch2024}. However, the impact of Pop. III stars and mini-halos on the EoR is unclear. Pop. III stars are likely to be the dominant contribution to the total star formation rate density (SFRD) at $z > 15-20$ and, if their IMF is more top-heavy than the present day one, Pop. III could significantly contribute to the heating and the ionization of the IGM which determines the evolution and shape of the 21cm signal \citep[e.g.][]{Qin2020, Jones2022, Sartorio2023}. 

The 21cm signal represents our most promising tool to put constraints on the thermal state of the IGM during the Cosmic Dawn and EoR. Even though no confirmed detection has been reported so far, the first upper limits on the 21cm power spectrum obtained with HERA phase I strongly disfavour cold reionization scenarios (\citealt{Hera2023}). % With HERA phase II we expect to either detect or rule out more reionization models. 
%{\textbf{[Next sentence still to see if we are actually gonna do it.]}} Another future tool that will be available once SKA-low becomes operative is directly imaging HII bubbles surrounding the first galaxies during the early stages of the EoR $(z \geq 9)$ \citep[e.g.][]{Geil2017, Mangena2020, Giri2021}. While individual bubbles will be too small to be detected, it is possible to directly detect the EoR at z $\sim$ 10 by stacking redshifted 21-cm spectra centered on known galaxies \citealt{Geil2017}. \\
In the last few years, the impact of Pop. III stars on the 21cm signal has been studied using both analytical and semi-analytical models \citep[e.g.][]{Cohen2017, Chatterjee2020, Mebane2020}. However these models either did not compute reionization (e.g. \citealt{magg2022, Hegde2023, Hector2024}), focusing only on the absorption trough of the 21cm signal occurring at $z \sim 13-20$, or used a very simple analytical approach to compute reionization (e.g.\citealt{Ventura2023}). On the other hand, \citet{Cohen2017}, \citet{Qin2021} and \citet{munoz2022} used a simple analytical model for modeling Pop. III star formation but computed the reionization self-consistently. 

In this work, we overcome these challenges using a realistic Pop. III star formation and mini-halo model (\citealt{Ventura2024}) developed within the semi-analytical model \textsc{meraxes} designed to self-consistently couple galaxy formation and reionization. While in \citet{Ventura2024} we ran this model on a small (L = 10 h$^{-1}$ cMpc) and high-resolution box, here we extend it to a significantly larger volume simulation (L = 210 h$^{-1}$ cMpc) enabling the study of cosmic reionization. Since at such large volumes we cannot directly resolve mini-halos, we implemented scaling relations between the SFRD and the dark matter density field calibrated on the results from the small and high-resolution box discussed in \citet{Ventura2024}.
%In this work we implemented Pop. III star formation and mini-halo physics in \textsc{meraxes} using a ($\sim 310$ cMpc)$^3$ volume. 
%Even though at such large volumes we cannot directly resolve mini-halos, we effectively implemented scaling relations between the SFRD and the dark matter density field calibrated on a smaller and higher resolution box whose results were discussed in \citet{Ventura2024}. 
With this new model we are able to accurately follow the evolution of the radiative backgrounds relevant to the EoR and 21cm signal (X-rays, Lyman-$\alpha$, ionizing UV, Lyman-Werner) and to disentangle the contribution of Pop. III star formation to the 21cm global signal and power spectrum. Pop. III stars are expected to have a stronger impact at $z \geq 15$ where they dominate star formation and ionization. Differently from previous works who explored the differences in the 21cm signal at Cosmic Dawn due to various Pop. III models, here we focus our attention on the residual signature of Pop. III on the 21cm power spectrum at $z \leq 10$ where the sensitivity of SKA is expected to be significantly better and a detection is more plausible. To achieve this, it is crucial to model both Pop. III star formation and reionization in a self-consistent framework. This study allows us to assess under which conditions an early heating of the IGM from Pop. III stars leaves a detectable imprint on the 21cm power spectrum at $z \leq 10$. 

This paper is structured as follows. In Section \ref{sec:modeling} we give a brief overview of Pop. III star formation in \textsc{meraxes}. In Section \ref{sec:scalingrels} we present the scaling relation between the SFRD in mini-halos and the density field calibrated from the small and high-resolution box which is implemented in the large (210 h$^{-1}$ cMpc)$^3$ box. In Section \ref{sec:21cmPS} we discuss the impact of different Pop. III star formation models on the 21cm power spectrum and in Section \ref{sec:21cmSense} we make forecasts on the observability of these power spectra with SKA. 
Finally, we summarize our main results and conclusions in Section \ref{sec:conclusions}. Our simulations use the best-fit parameters from the \citet{Planck16}: h = 0.6751, $\Omega_m$ = 0.3121, $\Omega_b$ = 0.0490, $\Omega_\Lambda$ = 0.6879, $\sigma_8$ = 0.8150, and n$_s$ = 0.9653.

\section{Pop. III galaxies in Meraxes}
\label{sec:modeling}

\textsc{meraxes}\footnote{\url{https://github.com/meraxes-devs/meraxes}} is a semi-analytical model (SAM) designed to study the interplay between galaxy formation and reionization (\citealt{Mutch2016, Qin2017, Qiu2019, Ventura2024}). \textsc{meraxes} includes a number of free parameters that are calibrated against observations (see Tables \ref{tab:GalParams} and \ref{tab:ReioParams}). Values in Table \ref{tab:GalParams} are calibrated against observed luminosity functions and stellar mass functions at $z \sim 5-8$ while those in Table \ref{tab:ReioParams} are calibrated against constraints on the neutral hydrogen fraction, ionizing emissivity and the Thomson scattering optical depth $\tau_e$ from \citet{Planck2020}. The most recent version of \textsc{meraxes} (\citealt{Ventura2024}, V24) includes Pop. III star formation and mini-halo physics. As shown in V24, the Pop. III parameters with the largest impact are the star formation efficiency $\alpha_{\rm SF, III}$ and the shape of the IMF. The latter has a large impact on galaxy evolution as it determines the strength of the feedback and the emission properties of the Pop. III stellar population. In the following sections we quickly summarize the main features of \textsc{meraxes} relevant for this work. 
%In V24, the free parameters regulating Pop. III star formation were extensively discussed, however

\subsection{Galaxy formation}
\label{sec:Meraxes}

\textsc{meraxes} post-processes the output of an N-body dark matter only simulation, reading the spatial and physical information of dark matter halos and computing the baryonic physics of galaxy formation. In particular processes included are: \textit{(i)} gas infall onto dark matter halos, \textit{(ii)} radiative cooling of the infalling gas, \textit{(iii)} star formation and \textit{(iv)} supernova and AGN feedback.\\
In V24 the cooling prescriptions were updated to account for H$_2$ cooling (the main cooling channel in mini-halos) and a more detailed metal enrichment model to keep track of the metallicity of each gas reservoir in a halo (which is crucial to distinguish between Pop. III and Pop. II star formation episodes). We also account for the effects of both baryon-dark matter streaming velocity and H$_2$ photo-dissociation by the Lyman-Werner background which increases the minimum mass of a mini-halo capable of hosting stars \citep[e.g.][]{Schauer2021}. Our model accounts for spatial variations only for the LW background, while for the relative velocity we assume a mean value throughout the entire box. For this work, we updated \textsc{meraxes} by adding the effect of H$_2$ self-shielding which counteracts the H$_2$ photo-dissociation, increasing the Pop. III SFRD by up to one order of magnitude at $z \sim 10$ (see e.g. \citealt{Feathers2024}). We discuss the details of the implementation and impact of H$_2$ self-shielding in \textsc{meraxes} in the Appendix \ref{sec:AppendixA}. \\
We refer the reader to \citet{Mutch2016} for a more detailed explanation of the main architecture of \textsc{meraxes}, \citet{Qiu2019} for the supernova model and V24 for the mini-halo model.\\
The main free parameters that regulate the galaxy formation in \textsc{meraxes} are summarized in Table \ref{tab:GalParams}. The Pop. II related ones are taken from \citet{Balu2023} where \textsc{meraxes} was run on a cosmological volume of L = 210 h$^{-1}$ cMpc resolving all atomic cooling halos and calibrated in order to match the observed UV luminosity functions at $z \sim 4-7$ (the agreement holds up to $z \sim 13$ as shown in \citealt{Qin2023}) and the stellar mass functions at $z \sim 5-8$. Given the lack of observations of Pop. III stars, the Pop. III parameters are largely unconstrained. The fiducial values adopted in this work are taken from V24 and their values are suggested from hydrodynamical simulations \citep[e.g.][]{Chon2021}.

\begin{table*}
    \centering
    \caption{Main free parameters for galaxy formation.}
    \label{tab:GalParams}
    \begin{tabular}{cccccccccc}
        \hline
        Parameter & Description & Fiducial value \\
        \hline
        $\alpha_{\rm SF, II}$ & Pop. II Star formation efficiency & 0.1 \\
        $\alpha_{\rm SF, III}$ & Pop. III Star formation efficiency & see Table \ref{tab:PopIIImodels} \\
        $\eta_0$ & Mass loading normalization & 7.0 \\
        $\epsilon_0$ & Supernova energy coupling normalization & 1.5 \\
        Z$_{\rm crit}$ & Critical metallicity for Pop III star formation & $10^{-4}$Z$_{\odot}$ \\
        $\Sigma_{\rm norm}$ & Critical surface density of cold gas for star formation & 0.37 M$_\odot$ pc$^{-2}$ \\
        Pop. III IMF & Shape of Pop. III IMF & Sal [1, 500] M$_\odot$\\
        E$_{\rm PISN}$ & Energy from Pair Instability SN & 10$^{52}$ erg\\
        E$_{\rm CCSN}$ & Energy from Core Collapse SN & 10$^{51}$ erg\\
        \hline
    \end{tabular}
\end{table*}

\begin{table*}
    \centering
    \caption{Main free parameters for reionization.}
    \label{tab:ReioParams}
    \begin{tabular}{cccccccccc}
        \hline
        Parameter & Description & Fiducial value \\
        \hline
        $f_{\rm esc, III}^0$ & Pop. III escape fraction normalisation & 0.14 \\
        $f_{\rm esc, II}^0$  & as above for Pop. II & 0.14 \\
        $\alpha_{\rm esc, III}$ & Pop. III escape fraction redshift scaling & 0.2 \\
        $\alpha_{\rm esc, II}$ & as above for Pop. II & 0.2 \\
        $L_{\rm X<2keV, III}/$SFR & specific Pop. III X-ray luminosity per unit star formation & see Table \ref{tab:PopIIImodels} \\
        $L_{\rm X<2keV, II}/$SFR & as above for Pop. II & 3.16 $\times 10^{40}$ erg s$^{-1}$ M$_\odot^{-1}$yr \\
        \hline
    \end{tabular}
\end{table*}

\subsection{Reionization and radiative backgrounds}

Together with galaxy formation, \textsc{meraxes} self-consistently computes the reionization and thermal evolution of the IGM using a modified version of the semi-numerical code 21cm\textsc{fast} (\citealt{Mesinger2011}). In this work, we compute the backgrounds relevant for the computation of the 21cm signal: the UV ionizing, X-rays, Lyman-$\alpha$ and Lyman-Werner (LW). The first is crucial to study the evolution of the reionization, while the X-ray and the Lyman-$\alpha$ backgrounds determine the thermal state of the IGM. In particular the X-ray background is likely to be the dominant contribution to the heating of the IGM once the first galaxies form and the lyman-$\alpha$ background is responsible for the coupling between the kinetic and the spin temperature of the neutral hydrogen. The LW background does not directly affect the IGM temperature, but determines whether or not mini-halos have enough molecular hydrogen to cool the gas and form Pop. III stars. Hereafter, we briefly summarize the key quantities that determine the evolution of these backgrounds. For a more detailed explanation on the implementations of these backgrounds, we refer the reader to \citet{Balu2023} for the UV, Lyman-$\alpha$ and X-ray and to V24 for the LW. \\ 
The ionizing background is mostly dependent on the SFRD, the average number of ionizing photons per stellar baryon N$_\gamma$ and the escape fraction of the UV photons f$_{\rm esc}$. The second quantity is mostly determined by the IMF: for Pop. II stars we adopt a Kroupa IMF which leads to $N_\gamma \sim 6000$. Since the Pop. III IMF is a free parameter in our model, $N_{\gamma, {\rm III}}$ is computed from the IMF adopted using the Pop. III stellar spectra from (\citealt{Raiter2010}; $N_{\gamma, {\rm III}} \sim 20000 - 70000$). f$_{\rm esc}$ is tuned to reproduce the EoR histories in agreement with observations. As per \citet{Balu2023}, we adopt a redshift-dependent escape fraction defined as followed:
\begin{equation}
    f_{\rm esc} = f_{\rm esc}^0\bigg(\frac{1+z}{6}\bigg)^{\alpha_{\rm esc}}.
\end{equation}
X-ray emission is mostly associated with high mass X-ray binaries (HMXB) and its contribution is proportional to the SFRD. In this work we use the widely adopted approximation for the comoving X-ray specific emissivity (erg s$^{-1}$ Mpc$^{-3}$) $\epsilon_x \propto$ L$_X /$ SFR $\times$ SFRD. Finally, we need to account for the fact that only photons with an energy below 2 keV (soft X-rays) are able to heat the IGM. As a result the main free parameter that regulates X-ray emissivity is the soft X-ray luminosity per unit star formation L$_{\rm X<2keV}$ / SFR. For Pop. II stars, this quantity is estimated from theoretical studies of emission spectra of HMXBs in low metallicity environments \citep[e.g.][]{Fragos2013, Madau2017, Das2017, Qin2020,Kaur2022}. For Pop. III stars there are no observational constraints as this quantity depends on the unknown Pop. III IMF. Recently, \citet{Sartorio2023} estimated the L$_X$ / SFR for Pop. III stars and found that for more top-heavy IMFs this quantity can be up to two orders of magnitude higher than the Pop. II value.

We highlight that in this work we separately compute the backgrounds from Pop. III and Pop. II stars due to the different spectra, properties and star formation rate density of the two distinct populations. 

\subsection{21cm physics}
\label{sec:21cmTheory}

Using the radiative backgrounds computed from the galaxy population in \textsc{meraxes}, we can estimate the 21cm signal. We encourage the reader to see \citet{FurlanettoRev}, \citet{Morales2010}, \citet{PritchardRev} and \citet{LiuRev} for reviews on the topic. Hereafter we only summarize the key equations used in this work (see also \citealt{Balu2023}). \\
We start with the 21cm brightness temperature field ($\delta T_{\rm b}$) which measures the deviation of the spin temperature of the neutral hydrogen (T$_S$) from the cosmic background T$_\gamma$ (i.e. the CMB). This is given by \citep{FurlanettoRev}:

\begin{equation}\begin{split}
    \delta T_{\rm b} &= \frac{T_S - T_\gamma}{1 + z}(1 - e^{-\tau_{\nu_0}}) \\
    &\simeq 27x_{\rm HI}(1+\delta_{\rm nl})\bigg(\frac{H}{dv_r/dr + H}\bigg)\bigg(1 - \frac{T_\gamma}{T_S}\bigg)\\
    &\times\bigg(\frac{1+z}{10}\frac{0.15}{\Omega_mh^2}\bigg)\bigg(\frac{\Omega_bh^2}{0.023}\bigg) \rm mK
    \label{eq:21cmTb},
\end{split}\end{equation}
where $\tau_{\nu_0}$ is the optical depth at the 21cm transition frequency $\nu_0$, x$_{\rm H}$ is the neutral hydrogen fraction, 1 + $\delta_{nl}$ is the density contrast in the dark matter field, $H(z)$ is the Hubble parameter at the redshift $z$, and $dv_r /dr$ is the radial derivative of the line-of-sight component of the
peculiar velocity. Once the cosmological model (\citealt{Planck16}) and the velocity and density field (from the N-body simulation) are fixed, $\delta T_{\rm b}$ is determined by the ionization and the spin temperature fields. The latter quantifies the population ratio of the two H$_{\rm I}$ hyperfine energy levels and is sensitive to the thermal state (i.e. the kinetic temperature T$_K$) of the gas as follows:

\begin{equation}
    T_S^{-1} = \frac{T_\gamma^{-1} + x_\alpha T_\alpha^{-1} + x_cT_{\rm K}^{-1}}{x_\alpha + x_{\rm c} + 1}
    \label{eq:Ts},
\end{equation}

where T$_\alpha$ is the colour temperature which we take equal to T$_{\rm K}$ while x$_\alpha$ and x$_c$ are the Lyman-$\alpha$ and collisional coupling coefficients, respectively. These coefficients quantify the strength of the processes (resonant scattering of Lyman-$\alpha$ photons, \citealt{Wouthuysen1952}, and collisions with free electrons) that drive the spin temperature towards the kinetic temperature (when x$_\alpha$ + x$_c$ $>>$ 1, T$_S \sim$ T$_K$ otherwise T$_S \sim$T$_\gamma$). T$_K$ is sensitive to the adiabatic cooling and to all the processes able to heat up (or cool) the IGM with the most dominant coming from the X-ray emission. Hence in this work we will consider only the X-ray heating neglecting the other source of heating such as primordial magnetic fields (\citealt{Minoda2019,Bera2020,Cruz2024a}), Lyman-$\alpha$ (\citealt{Ciardi2010,Reis2021,Mittal2021}), shocks (\citealt{Furlanetto2004,Gnedin2004,Ma2021}), cosmic rays (\citealt{Bera2023}), early accreting black holes (\citealt{Mebane2020, Ventura2023}) and decaying or annihilating dark matter (\citealt{Liu2018, Sun2023, Hou2024, Facchinetti2024}). Ultimately, the evolution of T$_S$ during the Cosmic Dawn and the EoR is mostly determined by the Lyman-$\alpha$ (for the coupling between T$_S$ and T$_K$) and X-ray flux.\\
Using Eq. \ref{eq:21cmTb} we can estimate both the all-sky averaged global signal and its fluctuations (i.e. the power spectrum). In this work we will often use the reduced power spectrum $\Delta_{21}^2$(k) = k$^3$/2$\pi^2$P$_{21}$(k) unless otherwise stated.
 
\section{Putting Pop. III galaxies and mini-halos in a large-scale simulation}
\label{sec:scalingrels}

In this section, we present a novel approach that enables us to efficiently estimate the SFRD from mini-halos in a large box (that does not directly resolve these objects) using results from a small, high-resolution box. %Differently from what has been done by \citet{Hazlett2024} who calibrated the \textit{Reinassance} simulation in order to account for Pop. III star formation by adding the previous star formation history to each atomic cooling galaxy resolved in the simulation, the methodology described in this section allows us to estimate only the \textit{total} SFRD occurring in mini-halos within a certain pixel of the simulation.  % Given that the aim of our work is to compute the impact of mini-halos to the thermal and reionization histories this is not a limitation \\
In Table \ref{tab:SimParams} we summarize the key parameters for both the small (L10) and large (L210) box.

\begin{table}
    %\centering
    \caption{Simulation parameters.}
    \label{tab:SimParams}
    \begin{tabular}{@{\extracolsep{\fill}}cccccc}
        \hline
        Label & Box side (cMpc) & Mass resolution (M$_\odot$) & Pixel side \\
        \hline
        L10 & 10 h$^{-1}$ & 4.71 $\times 10^5$ & 0.2 h$^{-1}$ \\
        L210 & 210 h$^{-1}$ & 3.16 $\times 10^7$ & 0.2 h$^{-1}$ \\
        \hline
    \end{tabular}
\end{table}

\subsection{Calibrating scaling relations from the L10 box}
Our starting point is the small ($L = 10 h^{-1}$ cMpc) high-resolution (halo mass resolution of M $\sim 4.7\times 10^5$ M$_\odot$) simulation used in V24. When building a scaling relation between SFR and other physical quantity, the first obvious choice is the dark matter density field $\delta$. For instance \citet{Munoz2023} showed, as a first-order approximation, SFR scales as e$^{\delta_R}$ where $\delta_R$ is the density field smoothed over a certain radius R and this relationship works quite well for $\delta \sim 0$ and large R ($\geq$ 3Mpc). 
\begin{figure*}
    \centering
    \includegraphics[width=\textwidth]{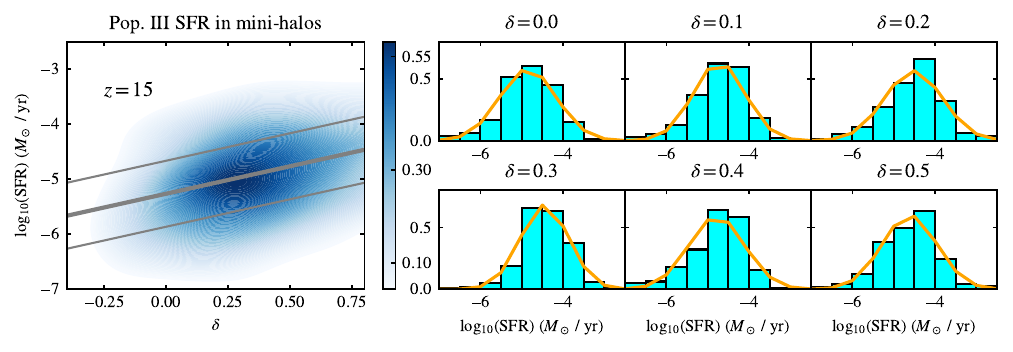}
    \caption{Left panel shows the density distribution of Pop. III star formation rate in mini-halos (M$_\odot$ yr$^{-1}$ in logarithm scale) vs the dark matter overdensity $\delta$ for each pixel at $z = 15$. The thick grey line shows the analytical fit SFR $\propto$ e$^{\delta}$ similar to the one adopted by \citet{Munoz2023} together with the 1$\sigma$ deviation (thin lines). For different values of $\delta$ (highlighted with the black rectangles) we show the distribution of Pop. III SFR in mini-halos (in logarithm scale) together with the best Gaussian fit.}
    \label{fig:SFRIIIfit}
\end{figure*}
To link our SFR in mini-halos with the density field, we first compute density, $\delta$({\textbf{x}}, $z$), and SFR grids for both Pop. III, SFR$_{\rm MC, III}$({\textbf{x}}, $z$), and Pop. II, SFR$_{\rm MC, II}$({\textbf{x}}, $z$), in the L10 box using the same grid resolution used in \citet{Balu2023} to compute reionization (L$_{\rm pixel} \sim 0.3$ cMpc) and accounting for the SFR within mini-halos. We highlight that our L$_{\rm pixel}$ is quite small compared to the smoothing radius R adopted by \citet{Munoz2023}, hence we expect a significant scatter in the above relation. We also split the contribution between Pop. III and Pop. II stars (a chemically enriched mini-halo will form Pop. II stars). In the left panel of Fig. \ref{fig:SFRIIIfit} we show our Pop. III SFR distribution as a function of the overdensity $\delta$ and with the grey line we highlight the SFR $\propto {\rm e}^{\delta_R}$ relation as in \citet{Munoz2023}. Despite the significant scatter for the reasons outlined above ($\sigma \sim 0.65$), the analytical approximation agrees with our results. The results shown hereafter are obtained from our fiducial simulation in V24.

\begin{figure}
    %\centering
    \includegraphics[width=\columnwidth]{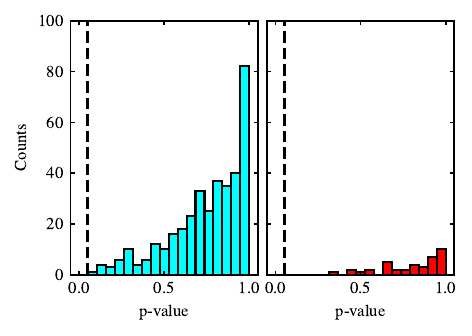}
    \caption{P-value distribution of K-S tests conducted on the Pop. III (left) and Pop. II (right) star forming pixels. Vertical line highlights the significant level of 0.05.}
    \label{fig:pVals}
\end{figure}

We start by investigating the distribution function of log$_{10}$(SFR) at a fixed overdensity $\delta$ and redshift log$_{10}$(SFR($\delta$, $z$)) finding that it follows a Gaussian distribution (or lognormal in the linear space).%\footnote{This is expected as fluctuations between pixels with the same density are fully uncorrelated. When looking at single galaxies this is not true anymore as there is a time correlation between different star formation episodes. This is why we apply this formalism to pixels (that contain several galaxies) and not to single galaxies.} 
We show results for log$_{10}$(SFR$_{\rm III}$) and selected values of $\delta$ in the smaller panels in Fig. \ref{fig:SFRIIIfit}. Hence, we can write:
\begin{equation}
    \Phi({\rm log_{10}(SFR_{MC}} | \delta, z)) = A(\delta, z)e^{\frac{({\rm log_{10}(SFR)} - {\rm log_{10}}(\overline{{\rm SFR}}(\delta, z)))^2}{2\sigma(\delta, z)^2}}
\end{equation}
where the normalization $A$, the mean log$_{10}(\overline{{\rm {SFR}}})$ and the standard deviation $\sigma$ all depend on the overdensity and redshift. The normalization is defined as the ratio between the number of SF pixels and the total number of pixels. We found the best fit parameters for each $\delta$ (grouped in bins of width = 0.1) and snapshot of the simulation. We tested whether the log$_{10}$(SFR) distribution function is indeed Gaussian by conducting a K-S test. P-values are calculated for each cell with SFR > 0 and taking a predefined significance level of 0.05 below which the null hypothesis will be rejected. Results are shown in Fig. \ref{fig:pVals} for both SFR$_{\rm{III}}$ and SFR$_{\rm{II}}$. P-values always exceed the significance level for both Pop. III (left panel) and Pop. II (right panel) SFR suggesting that the Gaussian distribution reproduces $\Phi({\rm log_{10}(SFR_{MC}} | \delta, z))$ both in the Pop. III and Pop. II cases. As expected, we see that there are far more Pop. III star forming pixels than Pop. II ones as mini-halos are more likely to form Pop. III stars. 

The next step is to study how the mean, standard deviation and normalization evolve with $\delta$ and $z$. 
\begin{figure}
    %\centering
    \includegraphics[width=\columnwidth]{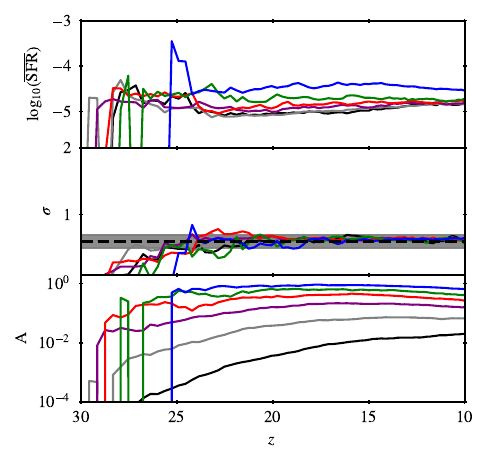}
    \caption{Redshift evolution of log$_{10}(\overline{{\rm SFR}})$ (top panel), $\sigma$ (mid) and A (bottom) for Pop. III for $\delta$ = 0.5 (black), 1.0 (grey), 1.5 (purple), 2.0 (red), 2.5 (green) and 3.0 (blue).}
    \label{fig:ParamIII}
\end{figure}
In Fig. \ref{fig:ParamIII} we show the redshift evolution of these parameters for $\delta = $ 0.5 (black), 1.0 (grey), 1.5 (purple), 2.0 (red), 2.5 (green) and 3.0 (blue). $\overline{{\rm SFR}}_{\rm MC, III}$ exhibits an almost constant trend in redshift and a correlation with $\delta$ (higher $\delta$ results in higher $\overline{{\rm SFR}}_{\rm MC, III}$). This demonstrates that $\overline{{\rm SFR}}_{\rm MC, III}$ is mostly determined by the number of Pop. III star forming halos in a pixel, which is higher for more overdense regions. Since Pop. III star formation episodes in mini-halos are often the first episode of star formation experienced by a galaxy, it is not impacted by supernova feedback\footnote{Even though supernova feedback can be neglected, this is not true for the Lyman-Werner background that halts star formation in mini-halos. For this reason we add a further condition that if a pixel is irradiated by a LW flux J$_{\rm LW} \geq$ J$_{\rm crit}$ above a critical threshold defined as M$_{\rm crit, MC} =$ M$_{\rm ato}$ (the atomic cooling threshold), $\Phi$(log$_{10}$(SFR$_{\rm MC})$) = 0. M$_{\rm ato}$ is the virial mass correspondent to a halo with a virial temperature T$_{\rm vir} = 10^4$ K. while M$_{\rm crit, MC}$ is defined in Eq. \ref{eq:MCritMC}.} so the Pop. III SFR is almost constant at all redshift.
This also explains why $\sigma$ is constant for all $z$ and $\delta$ ($\sigma_{{\rm MC, III}} \sim 0.65$). The parameter that is more sensitive to both $\delta$ and $z$ is the normalization. For very overdense regions ($\delta \geq 2$) it is almost one, meaning that almost all the overdense pixels host Pop. III SF mini-halos. For lower $\delta$ there is also an evolution in $z$ as, with cosmic time, lower density regions will host a larger number of mini-halos above the minimum mass for star formation. We repeated the same analysis for Pop. II star forming pixels (see Fig. \ref{fig:ParamII}). In this case the evolution is more scattered as Pop. II star formation episodes are not the first star forming episodes within a galaxy and so will be affected by both mechanical and chemical feedback from the previous history of the galaxy. This is also demonstrated by the larger standard deviation ($\sigma_{{\rm MC, II}} \sim 0.8$). The average value of $\overline{{\rm SFR}}_{\rm MC, II}$ is $\sim$ 1 order of magnitude higher than for Pop. III. This reflects the higher Pop. II star formation efficiency. Finally, A$_{{\rm II}}$ is always smaller than A$_{{\rm III}}$ showing that it is less likely for a mini-halo to form Pop. II stars.\\

\begin{figure}
    %\centering
    \includegraphics[width=\columnwidth]{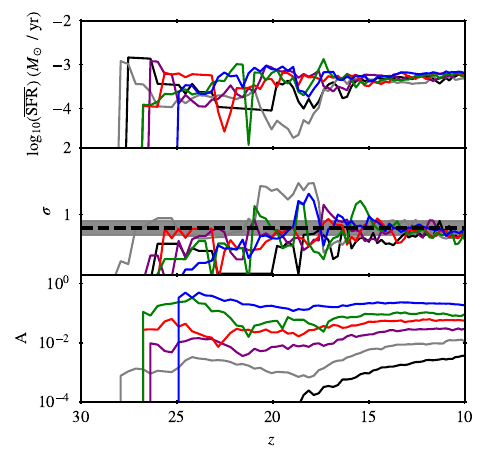}
    \caption{Same as Fig. \ref{fig:ParamIII} for Pop. II.}
    \label{fig:ParamII}
\end{figure}

\begin{figure}
    %\centering
    \includegraphics[width=\columnwidth]{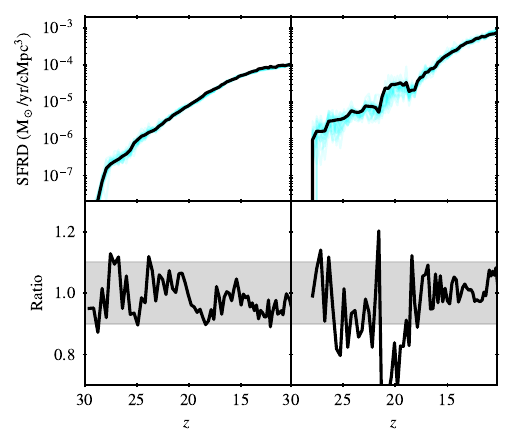}
    \caption{{\textbf{(top)}} Pop. III (left) and Pop. II (right) SFRD vs z from \textsc{meraxes} (black) and estimated from the density field (cyan). {\textbf{(bottom)}} ratio between the average of the 20 realizations for the SFRD estimated from the density field and the SFRD from \textsc{meraxes}.}
    \label{fig:SfrEstErr}
\end{figure}

We test this parametrization on the small box by estimating the mini-halo contribution to the Pop. III and Pop. II SFRD from the matter density field. To do this, we read the density grid at each $z$ and for each pixel assign a value of Pop. III SFR drawn randomly from $\Phi({\rm log_{10}( SFR_{MC, III}}))$ using $\delta$ of the pixel. We repeat the same procedure for SFR$_{\rm MC, II}$ adding the constraint that in order to have SFR$_{\rm MC, II} > 0$ that pixel needs to already have experienced a Pop. III star formation episode. This latter condition ensures a more realistic enrichment model (a pixel cannot have Pop. II star formation if it has not previously hosted Pop. III star formation). We ran twenty different realizations and for each realization estimated the SFRD$_{\rm MC, III}$ and SFRD$_{\rm MC, II}$ and compared with results of the simulations. Results are shown in Fig. \ref{fig:SfrEstErr}. In the upper panels we show the Pop. III (left) and Pop. II (right) SFRD from \textsc{meraxes} output (black line) and from each realization using the method outlined above (cyan shaded lines). In the lower panels we show the ratio between the average of the 20 realizations and the true SFRD from \textsc{meraxes}. All the realizations are in reasonable agreement with the data and the ratio is always lower than $\sim$ 10\% for both populations. This demonstrates the validity of the method outlined above. \\

The main advantage of this method is that it allows estimation of the SFRD from mini-halos in a simulation where these are not directly resolved. Parametrizing SFRD with a Gaussian distribution enables us to account for stochastic star formation (different pixels with same $\delta$ can have different SFR), without losing the correlation with the matter density field ($\overline{{\rm SFR}}, \sigma$ and A all depend on $\delta$). This method can be applied as long as the density field from both the low and the high resolution simulation share the same properties (mean, standard deviation) and it can be calibrated for any choice of parameters. 

\subsection{Applying scaling relations to the L210 box}

\begin{figure}
    %\centering
   \includegraphics[width=\columnwidth]{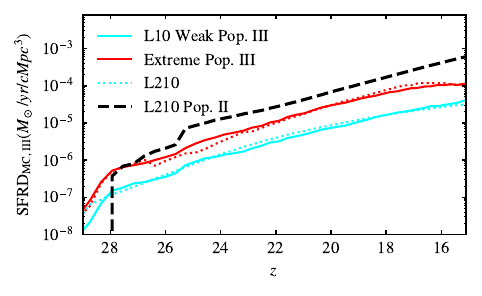}
    \caption{SFRD$_{\rm MC, III}$ vs $z$ from the L10 (solid) and L210 (dotted) box for two different Pop. III star formation models (see more details in text and Table \ref{tab:PopIIImodels}).}
    \label{fig:L10vsL210}
\end{figure}

We can now apply the methodology outlined in the previous section to the L210 box that can only resolve the atomic cooling halos by reading the density field and applying $\Phi({\rm log_{10}(SFR_{\rm MC}}))$ calibrated on different models.
In Fig. \ref{fig:L10vsL210} the mini-halo contribution to the Pop. III SFRD shows a good agreement between the two different simulations. 

\begin{figure*}
    \centering
    \includegraphics[width=\textwidth]{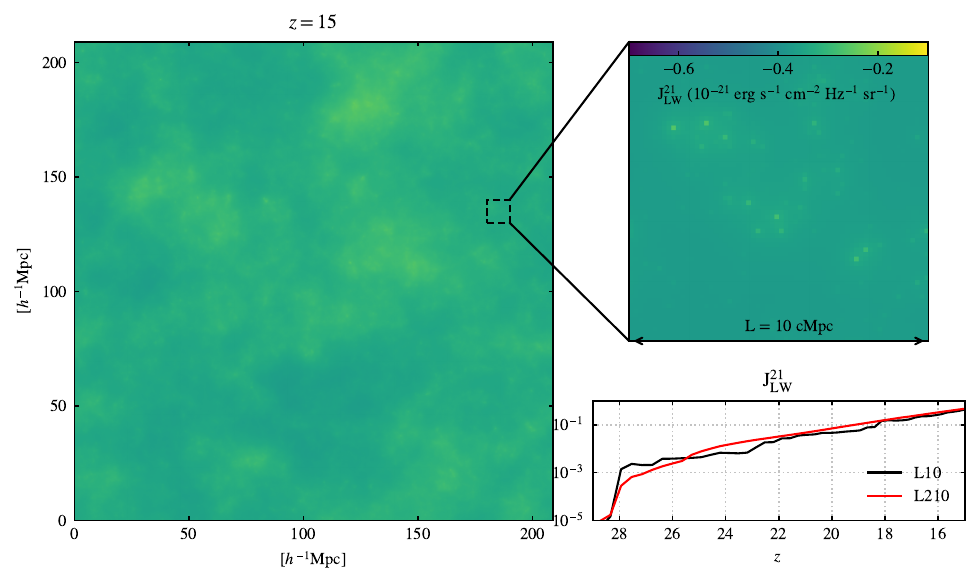}
    \caption{Left panel shows the 2D projections of the LW background (units of 10$^{-21}$ erg s$^{-1}$ cm$^{-2}$ Hz$^{-1}$ sr$^{-1}$ in the L210 box at $z=15$. Top right panel shows the same map but in the L10 simulation. Bottom right panel shows the redshift evolution of the average LW background (same units as above) in both the L10 (black line) and L210 (red line) simulation.}
    \label{fig:LWstats}
\end{figure*}

We highlight that our scaling relations do not explicitly depend on the Lyman-Werner background (except for the regions that are strongly irradiated by LW flux for which $\Phi$(log$_{10}$(SFR$_{\rm MC})) = 0$). This implicitly assumes that both the L10 and L210 box have similar LW backgrounds. We verify this assumption by computing the average LW background and LW maps in both simulations (See Fig. \ref{fig:LWstats} ). In the bottom right panel we show the redshift evolution of the LW background (in units of 10$^{-21}$ erg s$^{-1}$ cm$^{-2}$ Hz$^{-1}$ sr$^{-1}$) in the L10 (black line) and in the L210 box (red line). The two lines share similar trends showing that both simulations have a similar average LW background. The left and top right panels show the 2D projections of the LW field in the large and small box. These maps illustrate that the LW background is roughly uniform (as expected given that the mean free path of LW photons is $\sim 100$ Mpc). These plots demonstrate that the LW fields in the small and large box are indeed comparable showing that the estimation of the star formation in mini-halos with the scaling relations accounts for the radiative feedback\footnote{In this discussion we neglected the UV photo-ionizing feedback. This is justified by the fact that this effect impacts more the low-mass atomic cooling halos at $z \lesssim 10$ rather than mini-halos.}.  

Differently from what has been done by \citet{Hazlett2024} who calibrated a semi-analytical model to the \textit{Reinassance} simulation in order to account for Pop. III star formation by adding the previous star formation history to each atomic cooling galaxy resolved in the simulation, the methodology described in this section allows us to estimate only the \textit{total} SFRD occurring in mini-halos within a certain pixel of the simulation.
In this work we focus on three models of Pop. III star formation by varying three main parameters: the star formation efficiency, the IMF and the specific Pop. III X-ray luminosity per unit star formation while all the other free parameters (e.g. the escape fraction) are fixed at the fiducial value (see Tables \ref{tab:GalParams} and \ref{tab:ReioParams}). We chose to focus only on these three parameters since these have the strongest impact on both the evolution of the Pop. III SFRD and the amount of UV and X-ray photons emitted. The values chosen for the specific Pop. III X-ray luminosity per unit star formation are similar to those found in \citet{Sartorio2023} for the different IMFs explored in their work. Hereafter, we analyze three different Pop. III models designed to have the minimum, intermediate and maximum impact from Pop. III star formation in mini-halos, each model is summarized in Table \ref{tab:PopIIImodels}. The IMFs considered in this model are a Salpeter between 1 and 500 solar masses and a log-Normal IMF centered at 60 M$_\odot$ (see V24 for more details). We highlight that in our extreme Pop. III model we enhance star formation efficiency, L$_X$/SFR and the top-heaviness of the IMF at the same time. Each of these parameters has a different impact on the evolution of the 21cm signal (see Appendix \ref{sec:AppendixB} for a more detailed discussion of how each Pop. III parameter changes the evolution of the 21cm global signal and power spectrum).

\begin{table}
    %\centering
    \caption{Pop. III model parameters.}
    \label{tab:PopIIImodels}
    \begin{tabular}{@{\extracolsep{\fill}}cccccc}
        \hline
        Label & IMF type$^{\star}$ & $\alpha_{\rm SF, III}$ & $L_{\rm X<2keV, II}/$SFR \\
        \hline
        Weak Pop. III & Salpeter & 0.008 & $3\times10^{40}$\\
        Moderate Pop. III & Salpeter & 0.008 & $3\times10^{41}$\\
        Extreme Pop. III & logE & 0.08 & $3\times10^{42}$\\
        High SFE & Salpeter & 0.08 & $3\times10^{40}$ \\
        LogE & logE & 0.008 & $3\times10^{40}$ \\
        \hline
    \end{tabular}
    \\
    \footnotesize $^\star$see Table 2 in V24 for the details
\end{table}

\section{Impact of Pop. III star formation on 21cm physics}
\label{sec:21cmPS}

\begin{figure}
    %\centering
   \includegraphics[width=\columnwidth]{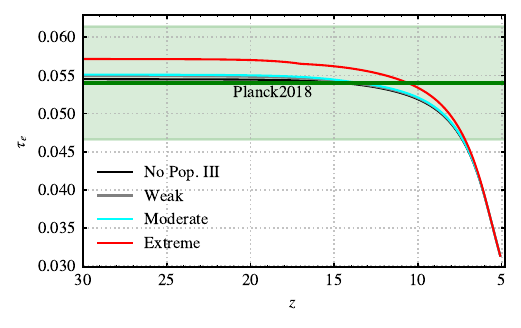}
    \caption{Integrated Thomson scattering optical depth $\tau_e$ computed for model with weak (grey), moderate (cyan), extreme (red) Pop. III and \citet{Balu2023} (black). The green curve and shaded region show the measurement of $\tau_e$ from the Planck 2018 collaboration (\citealt{Planck2020})}
    \label{fig:TauE}
\end{figure}

\begin{figure}
    %\centering
   \includegraphics[width=\columnwidth]{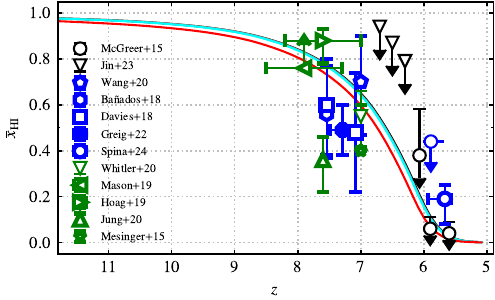}
    \caption{Constraints on the reionisation history (neutral hydrogen fraction vs z) for model with weak (grey), moderate (cyan), extreme (red) Pop. III and \citet{Balu2023} (black). The observational data are from analyses of dark pixels (\citealt{McGreer2015, Jin2023}), damping-wing absorption in quasar spectra (\citealt{Banados2018, Davies2018, Wang2020, Greig2022, Spina2024}) and equivalent width measurements (\citealt{Mesinger2015, Hoag2019, Mason2019, Jung2020, Whitler2020}) }
    \label{fig:xH}
\end{figure}

We can now estimate how different Pop. III star formation models in \textsc{meraxes} affect the 21cm signal. We started by verifying that, after introducing the additional Pop. III contribution to the fiducial Pop. II only model (\citealt{Balu2023}) we still obtain reionization histories consistent with the observational constraints on the Thomson scattering optical depth $\tau_e$ (Fig. \ref{fig:TauE}) and $\bar{x}_{\rm HI}$ (Fig. \ref{fig:xH}). The reionization histories from the models with Pop. III stars are only slightly modified and this negligible contribution comes from the secondary ionizations from X-rays. This is expected given that at z $\leq 15$ the Pop. III SFRD is at least one order of magnitude lower than Pop. II and their main contribution is expected from the X-ray emission rather than the UV.

\begin{figure}
    %\centering
   \includegraphics[width=\columnwidth]{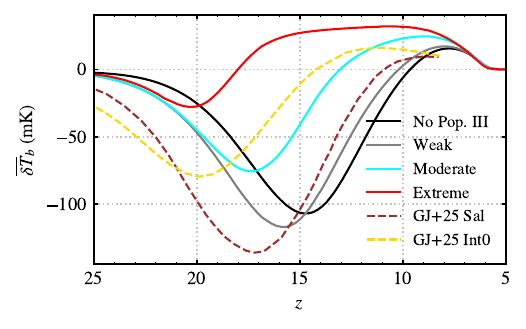}
    \caption{Effect of Pop. III star formation on the 21cm global signal ($\delta T_{\rm b}$ vs $z$). Pop. III models with small X-ray heating cause a stronger absorption at earlier times, while having a stronger Pop. III X-ray heating causes the signal to be seen in emission earlier. Color coding as in the previous figures. Brown and yellow dashed lines are taken from \citet{Jones2025} for a Salpeter (\textit{Sal}) and flat (\textit{Int-0}) IMF.}
    \label{fig:Tb}
\end{figure}

We compute the sky-averaged 21cm global signal (see Fig. \ref{fig:Tb}) without (black line) and with (grey, cyan and red line for weak, moderate and extreme Pop. III respectively) Pop. III star formation. As expected, introducing a Pop. III population with the same X-ray properties as Pop. II ones (i.e. weak Pop. III), simply shifts the absorption through to earlier epochs in virtue of the stronger coupling at higher-$z$ (see also \citealt{Ventura2023,Hegde2023}). However, if Pop. III stars have a stronger X-ray emission (i.e. moderate and extreme models) as suggested by \citet{Sartorio2023}, the absorption signal is quickly suppressed turning into an emission signal as early as $z \sim 18$ for the extreme Pop. III model and at  $z \sim$ 13 for the moderate Pop. III one. We note that a similar result has been found by a contemporaneous work by \citet{Jones2025} who found an analogous variation in the timing ($\Delta z \sim 3$) and depth ($\Delta \delta T_b \sim 50$ mK) of the absorption trough when considering a stronger X-ray contribution from Pop. III stars (in their model the L$_X$/SFR is self-consistently modeled from the IMF so that the difference between their \textit{Int-0} and \textit{Sal} model is of 2 orders of magnitudes.)

\begin{figure*}
    \centering
    \includegraphics[width=\textwidth]{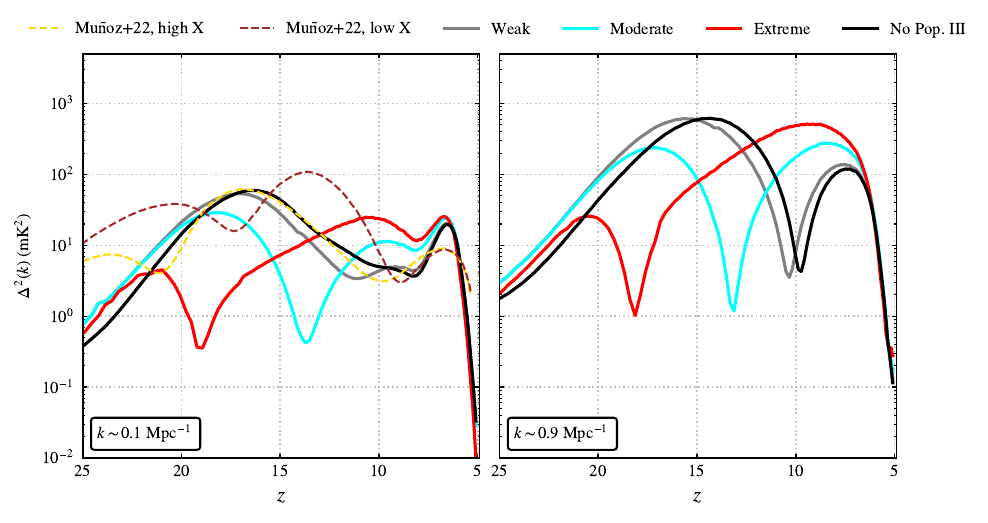}
    \caption{Effect of Pop. III star formation on the 21cm power spectrum ($\Delta {\rm _{21}}$ vs z) at large (k $\sim 0.1$ Mpc$^{-1}$) and small (k $\sim 0.9$ Mpc$^{-1}$) scales. Color coding as in the previous figures. Brown and yellow dashed lines are taken from \citet{munoz2022} for $k = 0.23$ Mpc$^{-1}$ assuming a weak and strong X-ray luminosity per unit of Pop. III SFR (bottom right panel of Fig. 17.)}
    \label{fig:Psvsz}
\end{figure*}

As shown in Fig. \ref{fig:Psvsz} earlier coupling and heating from Pop. III impacts the 21cm power spectrum both at the large and small scales. The models considered here produce 21cm signals that are different not only during the coupling and heating epoch ($z \sim 10 - 20$), but also at lower redshift when the reionization is in progress. The impact is stronger at smaller scales where models with a stronger heating exhibit a larger power spectrum at $z \sim 7 - 10$. This is of crucial importance as current and upcoming facilities are improving their observations at $z \leq$ 10 (see more discussion in Section \ref{sec:21cmSense}). 

We compare our results with the ones in \citet{munoz2022} who studied how different Pop. III parameters impact the 21cm signal using 21cm\textsc{Fast}. In the left panel of Fig. \ref{fig:Psvsz} the yellow and brown dashed lines are obtained by changing the specific Pop. III X-ray luminosity by a factor of 9. The global trends are similar with our models with stronger Pop. III X-ray emission (dashed yellow line and red/cyan lines) showing an earlier and weaker first peak compared to low X-ray models (dashed brown and grey lines). In their model, different Pop. III X-ray properties strongly impact the position and amplitude also of the second peak. In our models instead, the position of the second peak is only slightly anticipated in the strong X-ray models. This difference is likely due to the fact that our second peak occurs at much lower redshift ($z \sim 10$) when most of the emission comes from Pop. II stars. Finally, differently from \citet{munoz2022} at $z \sim 10$ simulations with a stronger Pop. III X-ray heating have a significantly larger power spectrum compared to weak X-ray models (a similar result has been found also by \citealt{Jones2025}). 

This analysis shows that Pop. III star formation not only impacts the 21cm global signal during the cosmic dawn as previously assessed \citep[e.g.][]{Qin2020, Jones2022, Ventura2023, Hegde2023, Hector2024} but also that an early ($z \geq 15$) heating of the IGM provided by this population leaves a strong signature on the power spectrum during the EoR ($z \leq 10$). The impact on the power spectrum is stronger for models that have a large Pop. III X-ray emissivity (i.e. moderate and extreme Pop. III models) while models with a milder X-ray emission (i.e. weak Pop. III) have a stronger effect on the global signal. Ultimately, this tells us that the power spectrum during the EoR can be used to disentangle different heating models and potentially constrain the properties of Pop. III stars. We highlight that here we considered only the X-ray emission from stellar remnants. While there might be other sources that significantly heat the IGM at $z \geq 15$ (see discussion at end of Section \ref{sec:21cmTheory}), the other effects are likely to be either subdominant, still related to Pop. III stars (e.g. cosmic rays) or dominant only at the dark-ages (e.g. dark matter annihilation).
%There might be other sources that significantly heat the IGM at $z \geq 15$ whose impact would add to the Pop. III one considered here.

Finally, we note that in this work we did not include the effect of the X-ray feedback on Pop. III star formation. As noted by \citet{Ricotti2016,Park2021}, X-rays have both a positive and negative effect on Pop. III star formation as they heat the gas and increase the electron fraction. The heating makes gas accretion more difficult (hence delaying star formation) and free electrons promote the formation of H$_2$ making the molecular cooling more efficient. In presence of a strong X-ray background this latter effect is dominant at $10 \lesssim z \lesssim 20 $ (see e.g. Fig. 9 in \citealt{Hegde2023}). Hence, including the X-ray feedback would likely increase the Pop. III SFRD making the impact of Pop. III stars even stronger than what predicted here especially for the moderate and extreme models.

%\subsection{Comparison with previous works}

%{\textbf{A few other groups have modeled Pop. III star formation in (semi-)analytical and/or numerical models in order to assess their effect on the 21cm signal. In }}

\section{Observability with SKA}
\label{sec:21cmSense}

\begin{figure}
    %\centering
    \includegraphics[width=\columnwidth]{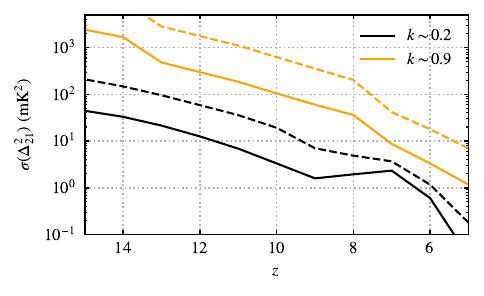}
    \caption{21cm power spectrum sensitivity as a function of redshift $k \sim 0.2$ (black line) and 0.9 cMpc$^{-1}$ (orange line) assuming a 1000 (solid lines) and 180 hours (dashed lines) observation with SKA.}
    \label{fig:SenseRaw}
\end{figure}

In the previous section we showed that an early heating of the IGM significantly affects the 21cm power spectrum during the EoR. Now we want to investigate how our models compare with the currently available upper limits and whether their differences will be observable with SKA. In this section we consider the power spectrum at $z \leq 10$ as the current and upcoming interferometers are significantly less sensitive at higher redshift. To better appreciate this, in Fig. \ref{fig:SenseRaw} we show the power spectrum noise (mK$^2$) with SKA as a function of redshift at $k \sim 0.2$ and 0.9 cMpc$^{-1}$ assuming a 1000 hours (solid lines) and 180 hours (dashed lines) observations with SKA. At higher redshift the noise steadily increases and already at $z \geq 10$ the noise is of the order of 10s to 100s mK$^2$ with 1000 hours observation. 

\begin{figure*}
    \centering
    \includegraphics[width=\textwidth]{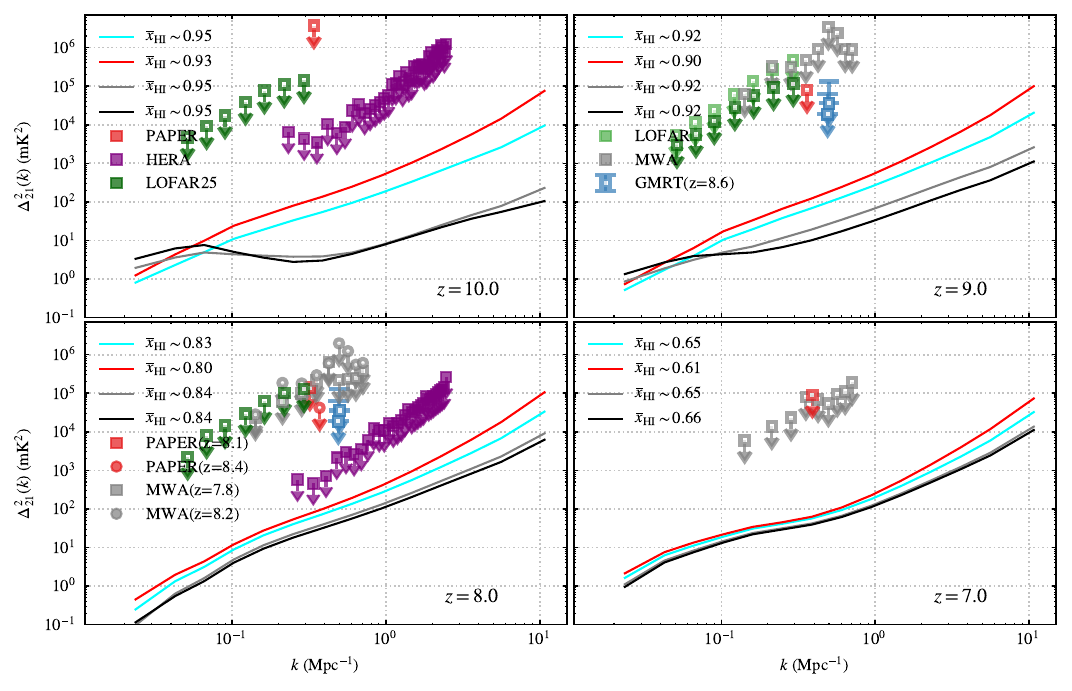}
    \caption{21cm power spectrum at $z$ = 10, 9, 8 and 7 as a function of $k$ for different Pop. III star formation models. Color coding as in the previous figures. We show also current upper limits from PAPER at $z = 9.9, 8.7, 8.4, 8.1, 7.5$ (\citealt{Kolopanis2019}), LOFAR at $z = 10.1, 9.1, 8.3$ (\citealt{Mertens2020, Mertens2025}), MWA at $z =8.7, 8.2, 7.8$(\citealt{Trott2020}), GMRT at $z = 8.6$ (\citealt{Paciga2013}) and HERA (\citealt{Hera2023}) at $z = 10, 8$.}
    \label{fig:Psvsk}
\end{figure*}

Firstly, we compare our predictions with current upper limits from a number of facilities including MWA, LOFAR, GMRT, PAPER and HERA (we focus at $z = 7 - 10$ where most of the measurements have been taken). 
%We focus on a number of facilities including MWA, LOFAR, GMRT, PAPER and HERA who already provide upper limits (we consider $z = 7 - 10$ where most of the measurements have been taken).
In Fig. \ref{fig:Psvsk} we show the 21cm power spectrum at $z$ = 10, 9, 8 and 7 for all the four models together with the available upper limits (see label). All our models are below these upper limits. However, the moderate and extreme Pop. III models are closer to the current HERA constraints at $z$ = 8 suggesting that soon these models can be either detected or ruled out. In our model the main effect of different Pop. III models is to change the timing and amplitude of the peaks in the 21cm power spectrum rather than introducing a specific feature. We note that we did not account for spatial variations in the velocity acoustic oscillations (see Section \ref{sec:Meraxes}) which would introduce wiggles in the $\Delta^2_{21}$ at scales $k \sim 0.1$ Mpc$^{-1}$ (\citealt{Hector2024}). While this effect can be important when focusing on the 21cm power spectrum during the coupling and heating epochs, these fluctuations are quickly washed out at $z \lesssim 13$ (see Fig. 13 and Section 6C in \citealt{Hector2024}).

We next consider the upcoming SKA. In order to estimate the observability of the four models analyzed so far, we perform a similar analysis as in \citet{Balu2023b} that we briefly summarize hereafter.
The sensitivity of a radio interferometer is mostly regulated by the thermal noise ($\Delta_{\rm N}$) and the cosmic variance with the former dominating the noise at small scales and the latter at large scales. The thermal noise is related to the bandwidths of the instrument, beam factor (see \citealt{Parsons2014}), the integration time of the mode $k$ and the temperature of the system (given as the sum of the sky and receiver temperature)  (\citealt{Morales2005,McQuinn2006,Parsons2012}). We can hence write the total noise by summing these two components:

\begin{equation}
    \bigg[\frac{1}{\sigma[\Delta^2_{21}(k)]}\bigg]^2 = \Sigma_i\bigg(\frac{1}{\Delta^2_{\rm N} + \Delta^2_{21}}\bigg)^2 .
    \label{eq:21cmSense}
\end{equation}
By doing so, we are effectively assuming that the errors are Gaussian distributed, which is reasonable for the relevant scales in this work (\citealt{Qin2021,Prelogovic2023}). Finally, a 21cm detection is heavily limited by the ability of removing the foregrounds. We used the python package \textsc{21cmsense}\footnote{\url{https://github.com/rasg-affiliates/21cmSense}} (\citealt{Pober2013,Pober2014}) which, given the specifics of an interferometer, a mock 21cm power spectrum and an observational campaign, computes the interferometer sensitivity to the 21cm power spectrum under different assumptions of foreground removals. We used the assumption 'moderate' foreground removals and we focused on the first phase of SKA (i.e. SKA1-low), in particular we included the stations in the 'Central Area' of the SKA1-low\footnote{See the official SKA1 System Baseline Design document in \url{ https://www.skao.int/en} for further details}, resulting in 296 stations of diameter 35m distributed across a circular area with 1.7 km diameter. We assumed two different observational campaigns: one of 180 hours (six hours per night for 30 days) and another of 1080 hours (six hours per night for 180 days). Additionally, we avoid the modes that are contained within the foreground wedge (\citealt{Datta2010}) so that effectively we are limited to $k_{\rm min}$ = 0.16 Mpc$^{-1}$ for our analysis and $k_{\rm max}$ = 1.4 Mpc$^{-1}$ (the latter arises from a combination of the spatial scales that are probed by the SKA1-low and the shot noise from our simulation).

To quantify the detectability of a Pop. III model we compute by how many sigma its power spectrum differs from the one arising from the No Pop. III model:
\begin{equation}
    D_i(k) = \frac{\Delta^2_{21,i}(k) - \Delta^2_{21,0}(k)}{\sqrt{\sigma^2[\Delta^2_{21,i}(k)]+ \sigma^2[\Delta^2_{21,0}(k)]}},
    \label{eq:Detectability}
\end{equation}
where the subscript $i$ refers to the Pop. III model considered and 0 to the No Pop. III model. To detect a Pop. III model the difference between its power spectrum and the No Pop. III one needs to be at least of 1$\sigma$ otherwise the model is not detectable.
\begin{figure}
    \centering
    \includegraphics[width=\columnwidth]{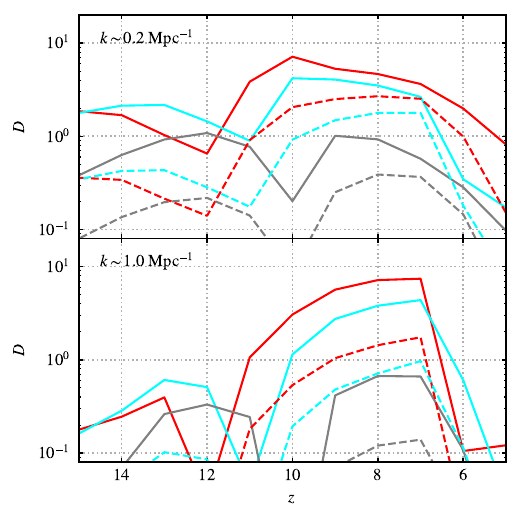}
    \caption{Detectability (see Eq. \ref{eq:Detectability}) vs $z$ for each Pop. III model (color coding as in the previous figures) at $k\sim 0.2$ (upper panel) and 1.0 Mpc$^{-1}$ (lower panel). Solid (dashed) lines assume a 1080 (180) hours observation with SKA1-low.}
    \label{fig:detectability}
\end{figure}
In Fig. \ref{fig:detectability} we show the evolution in redshift of this quantity at $k\sim 0.2$ (upper panel) and $k\sim 0.9$ Mpc$^{-1}$ (lower panel) for each Pop. III model and assuming 1080 (solid lines) and 180 (dashed lines) observation hours. The weak Pop. III model (grey line) is different from the No Pop. III one by less than 1$\sigma$ both at the small and large scales even with an observation of 1080 hours. The moderate Pop. III model is several $\sigma$ away from the No Pop. III model at $7\lesssim z \lesssim 10$ both at the small and large scales with a 1080 hours observation while with a 180 hours the model is detectable only at the large scales where the difference is between 1 and 2 $\sigma$. The extreme Pop. III model is the one with the largest difference from the No Pop. III model being potentially detectable at both large and small scales even with a 180 hours observations. 
% We show the power spectrum noise (shaded regions) generated for our 21cm power spectra for different Pop. III models assuming 180 and 1080 hours of observations in Fig. \ref{fig:Sense180} and \ref{fig:Sense1000}.

\begin{figure*}
    \centering
    \includegraphics[width=\textwidth]{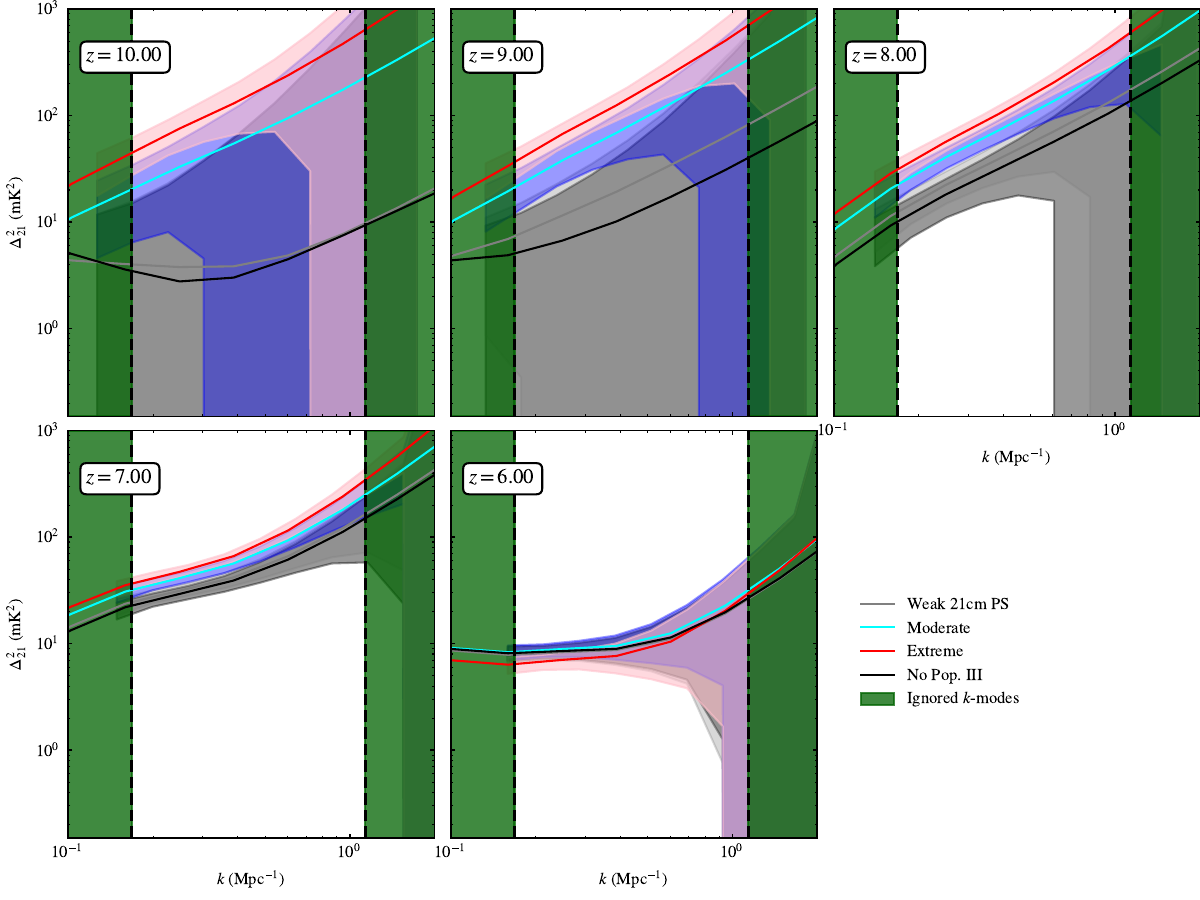}
    \caption{The black, grey, red and cyan curves show the 21cm power spectrum from all the models in Table \ref{tab:PopIIImodels}. The shaded region of the correspondent color represents the sensitivity (including both thermal and cosmic variance noise) to the associated  21cm power spectrum for a 180 hours observation with the upcoming SKA1-low with the 'moderate' foreground removal case. We ignore (green shaded regions) all the $k$-modes falling within the 21cm foreground wedge (see text for further details).}
    \label{fig:Sense180}
\end{figure*}

\begin{figure*}
    \centering
    \includegraphics[width=\textwidth]{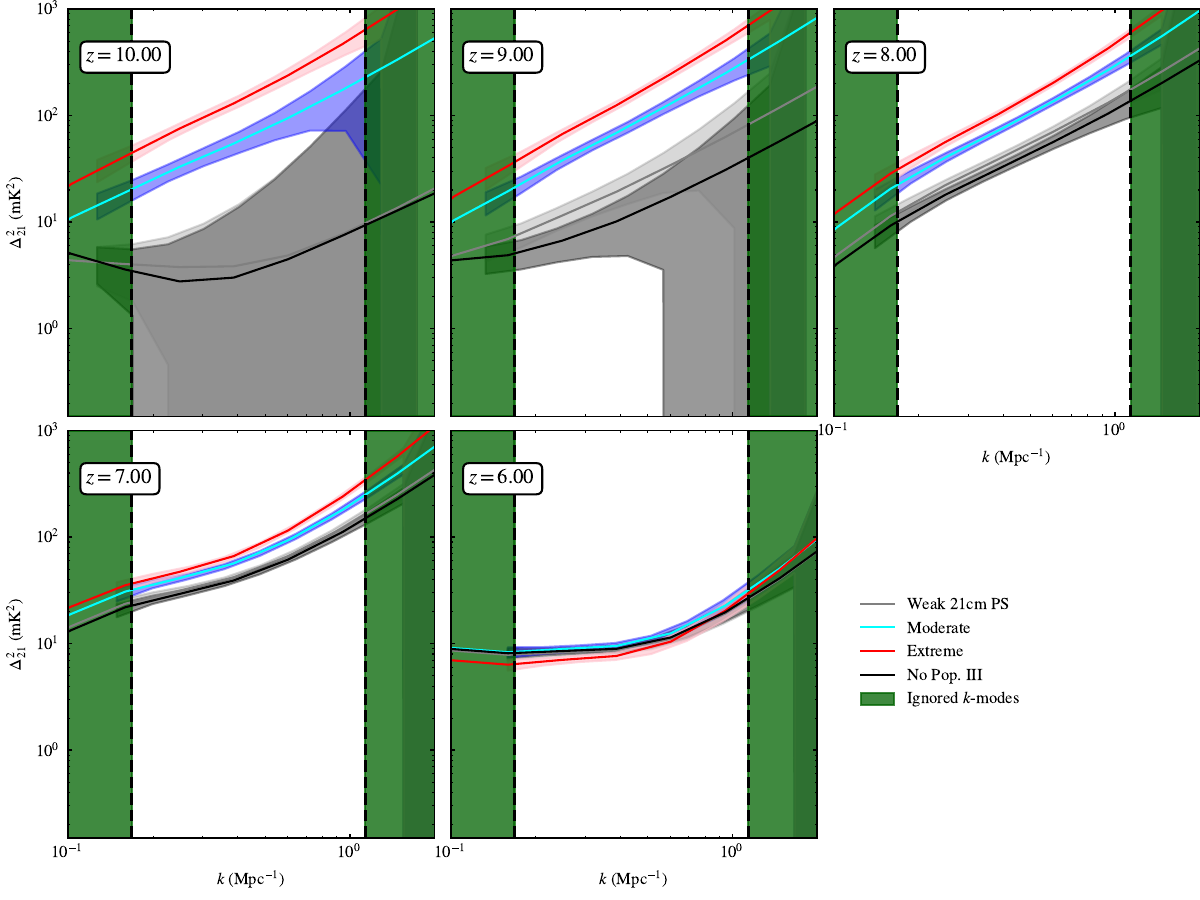}
    \caption{Same as Fig. \ref{fig:Sense180} but assuming 1080 hours observation.}
    \label{fig:Sense1000}
\end{figure*}

%With an observational campaign of 180 hours, all the models are essentially indistinguishable (the shaded regions superimpose to each other). However, 
Finally, we highlight the differences between each Pop. III model by showing their power spectrum noise (shaded regions) assuming 180 and 1080 hours of observations in Fig. \ref{fig:Sense180} and \ref{fig:Sense1000}.
Models with stronger X-ray contribution (red and cyan lines) have a power spectrum that differs from the models with no or mild Pop. III contribution (dark and light grey regions) by more than the observational uncertainty. The 1$\sigma$ region of its power spectrum (pink shaded region) is above the models with no or mild Pop. III contribution  at $z$ = 7 and 8 for all the $k$-modes, at $z$ = 9 at $k \lesssim$ 0.7 Mpc$^{-1}$ and at $z$ = 10  only at the largest scales considered ($k \lesssim$ 0.4 Mpc$^{-1}$). With only 180 hours observations however, it is not possible to clearly distinguish between the intermediate and extreme Pop. III model (blue an pink shaded regions). At $z$ = 6 it is not possible to distinguish any of these models since at this redshift the power spectrum is mostly determined by the ionization state of the IGM which is identical for all the four models considered. 

The sensitivity greatly improves for an observational campaign of 1000 hours (Fig. \ref{fig:Sense1000}). In this case, models accounting for Pop. III early heating which is stronger than Pop. II are clearly distinguishable from models with no or mild Pop. III contribution. At $z$ = 10, 9 and 8, it is even possible to distinguish between intermediate and extreme Pop. III models at all scales within $k_{\rm min}$ and $k_{\rm max}$, indicating that SKA observations of the 21cm power spectrum during the EoR are sensitive to the IGM heating occurred at $z \geq 15$. At $z$ = 7, the difference between the various models are less evident but intermediate and extreme Pop. III models still fall outside the uncertainty regions of weak Pop. III and No Pop. III. This indicates that models with a harder Pop. III X-ray emission can be disentangled from models with no or mild Pop. III X-ray emission even at $z$ = 7. Also \citet{Jones2025} focused on the possibility of detecting difference Pop. III models using 21cm observations. In their analysis they included both REACH (sensitive to the 21cm global signal) and SKA-Low confirming that an observation of $\sim 3000$ hours with SKA-Low can constrain the Pop. III IMF due to their difference in the X-ray heating. Finally, we highlight that the impact of Pop. III stars is visible on the power spectrum as long as their X-ray contribution is boosted compared to Pop. II. The differences between the weak Pop. III and the No Pop. III scenario are so small they will not be visible even with 1000 SKA hours. This demonstrates that an upcoming 21cm power spectrum detection will shed light on the properties of the first sources establishing their contribution to the heating of the IGM. 

%\begin{equation}
%    \Delta_{\rm N}^2(k) \simeq {\rm X}^2{\rm Y}\frac{k^3}{2\pi^2}\frac{\Omega'}{2t}{\rm T}^2_{\rm sys}
%\end{equation}

\section{Conclusions}
\label{sec:conclusions}

In this work we investigated how the early ($z \geq 15$) heating of the IGM due to Pop. III stars impacts the 21cm power spectrum during the EoR ($z \leq 10$) which will be observable by the SKA. We developed a framework of scaling relations between Pop. III star formation rate in mini-halos and the density field within the semi-analytical model \textsc{meraxes} that allows us to account for the radiative contribution of mini-halos that cannot be directly resolved in a large (L = 210 h$^{-1}$ cMpc) cosmological simulation. The scaling relations are calibrated on a smaller and higher resolution simulation able to resolve all the mini-halos. We then investigated three Pop. III models each with different IMF, star formation efficiency and specific X-ray luminosity. This formalism is both based on realistic galaxy population and it provides an accurate estimation of the Pop. III backgrounds yielding a computation of the ionization and thermal state of the IGM which is crucial to obtain a reliable estimation of the 21cm signal and power spectrum during the EoR ( $z \lesssim 10$). The key results can be summarized as follows:

\begin{enumerate}
    \item The inclusion of Pop. III stars does not significantly change the EoR history as their main contribution comes from the secondary ionizations from their X-ray emission. As a result, all the models that we considered give reionization histories consistent with the latest constraints on the neutral hydrogen fraction and the Thomson scattering optical depth.
    \item As previously known, the evolution of the 21cm global signal during the Cosmic Dawn is significantly impacted by Pop. III stars. In a model where Pop. III and Pop. II have the same X-ray specific luminosity (weak Pop. III) the absorption through of the 21cm global signal is slightly deeper and occurs at higher-z. If instead Pop. III stars have a larger L$_X$ / SFR, the signal is seen in emission at higher-$z$ ($z \sim 14$ for the intermediate Pop. III model and $z \sim 18$ for the extreme Pop. III model).
    \item The heating from Pop. III stars significantly increases the 21cm power spectrum during the EoR both at large and small scales. This is more evident for models with an increased specific X-ray luminosity where the difference can be of more than a factor of 4 at z $\gtrsim 7$. This shows that the 21cm power spectrum during the EoR is sensitive to the heating occurred at much higher redshift.
    \item Focusing on a possible future detection of the 21cm power spectrum, we estimated the observational uncertainty of each model using the \textsc{python} package \textsc{21cmsense} assuming an observational campaign with SKA1-low of 180 and 1080 hours respectively. We found that 180 hours are not enough to distinguish different Pop. III models but they already can confirm/exclude the presence of a stronger heating from Pop. III stars at high-$z$ since their power spectra differs by  more than 1$\sigma$ from the No Pop. III model. However, using $\sim$ 1000 hours of SKA, models with a stronger Pop. III contribution (stronger X-ray emission and/or higher Pop. III star formation efficiency) become clearly distinguishable at $z \gtrsim 7$.
    %as the separation between different models is larger than the 1$\sigma$ observational uncertainty. This does not apply toward the end of the EoR (z $\lesssim 6$) or for a model with the a lower Pop. III X-ray emissivity (weak Pop. III). In the former case this is because at z $\sim 6$ the power spectrum is mostly determined by the ionization state of the IGM (which is not impacted by Pop. III star formation in our models). When the Pop. III X-ray emissivity is the same as the Pop. II one, the 21cm power spectrum is only very mildly increased compared to the scenario without Pop. III stars. 
    The power spectrum towards the end of the EoR ($z \lesssim 6$) does not distinguish Pop. III models.
\end{enumerate}

\section*{Acknowledgements}

%Acknowledge computing time awarded? Acknowledge Nordita?
%Do I still acknowledge Astro3D?

We thank the anonymous referee for very constructive comments that help improve the quality of this paper.
This research was supported by the Australian Research Council Centre of Excellence for All Sky Astrophysics in 3 Dimensions (ASTRO 3D, project \#CE170100013). This work was performed on the OzSTAR national facility at Swinburne University of Technology. OzSTAR is funded by Swinburne University of Technology and the National Collaborative Research Infrastructure Strategy (NCRIS). This research was also undertaken with the assistance of resources from the National Computational Infrastructure (NCI Australia), an NCRIS enabled capability supported by the Australian Government. SB acknowledges support from grant PID2022-138855NB-C32 funded by MICIU/AEI/10.13039/501100011033 and ERDF/EU, and  project PPIT2024-31833, cofunded by EU--Ministerio de Hacienda y Función Pública--Fondos Europeos--Junta de Andalucía--Consejería de Universidad, Investigación e Innovación. YQ is supported by the ARC Discovery Early Career Researcher Award (DECRA) through fellowship \#DE240101129. Finally, EMV thank A. Mesinger and J. B. Munoz for insightful discussions regarding the development of the scaling relation model and C. Power for assistance with the \textit{N}-body simulation.

%%%%%%%%%%%%%%%%%%%%%%%%%%%%%%%%%%%%%%%%%%%%%%%%%%
\section*{Data Availability}

The main data presented in this work have been analysed with the publicly available  \textsc{python} package \textsc{dragons}\footnote{\href{https://github.com/meraxes-devs/dragons}{https://github.com/meraxes-devs/dragons}}. The latest version of \textsc{meraxes} is publicly available on GitHub. Further data and codes will be shared on reasonable request to the corresponding author.



\bibliographystyle{mnras}
\bibliography{meraxes_scaling} % if your bibtex file is called example.bib

%\begin{thebibliography}{99}
%\end{thebibliography}

%%%%%%%%%%%%%%%%%%%%%%%%%%%%%%%%%%%%%%%%%%%%%%%%%%

%%%%%%%%%%%%%%%%% APPENDICES %%%%%%%%%%%%%%%%%%%%%

\appendix

\section{Including the H$_2$ self-shielding in \textsc{meraxes}}
\label{sec:AppendixA}

\begin{figure}
    %\centering
    \includegraphics[width=\columnwidth]{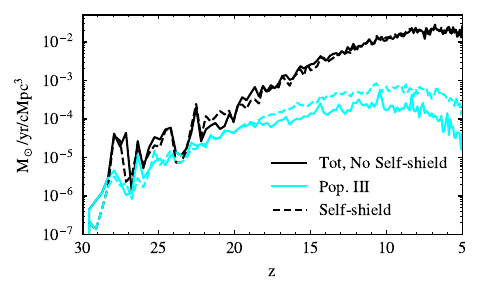}
    \caption{Total (black) and Pop. III (cyan) SFRD vs redshift without (solid) and with (dashed) accounting for H$_2$ self-shielding.}
    \label{fig:SFRD_SS}
\end{figure}

Many recent simulations \citep[e.g.][]{Nebrin2023,Hegde2023} have assessed the importance of incorporating H$_2$ self-shielding in mini-halos as it significantly changes the minimum mass of mini-halos capable of cooling down the gas via molecular cooling (M$_{\rm crit, MC}$). While this was not implemented in V24, for this work we now included it by updating M$_{\rm crit, MC}$ (Eq. 4 in V24) taking the fitting function found in \citet{Kulkarni2021}:
\begin{equation}
     M_{\rm crit, MC} = M_{z = 20}(J_{\rm LW}, v_{\rm bc})\bigg(\frac{1+z}{21}\bigg)^{-\alpha(J_{\rm LW}, v_{\rm bc})}
\end{equation}
where 
\begin{equation}\begin{split}
    M_{z=20}(J_{\rm LW}, v_{\rm bc}) & = (M_{z = 20})_0(1 + J_{\rm LW}/J_0)^{\beta_1} \\
    & \times (1 + v_{\rm bc} / v_0)^{\beta_2}(1 + J_{\rm LW}v_{\rm bc}/Jv_0)^{\beta_3}
    \label{eq:MCritMC}
\end{split}\end{equation}
and
\begin{equation}\begin{split}
    \alpha(J_{\rm LW}, v_{\rm bc}) & = \alpha_0(1+J_{\rm LW}/J_0)^{\gamma_1}(1+v_{\rm bc}/v_0)^{\gamma_2} \\ 
    & \times (1+J_{\rm LW}v_{\rm bc}/J_0v_0)^{\gamma_3}. 
\end{split}    
\end{equation}
M$_{\rm crit, MC}$ is a function of the Lyman-Werner flux (J$_{\rm LW}$) in $10^{-21}$ erg s$^{-1}$ cm$^{-2}$ and the rms streaming velocity at recombination (v$_{\rm bc}$) in km s$^{-1}$. This result is normalized to the minimum halo mass at $z = 20$ in absence of LW background and streaming velocity (M$_{z = 20})_0$ and to typical values for LW background and streaming velocity ($J_0 = 1, v_0 = 30$). $\alpha_n$ and $\beta_n$ are free parameters calibrated to fit results of hydrodynamical simulations in \citet{Kulkarni2021}. This fit accounts for the H$_2$ self-shielding, so it makes the effect of the LW background on M$_{\rm crit,MC}$ less strong compared to what was previously assumed in V24 (\citealt{Visbal2015}). As shown in Fig. \ref{fig:SFRD_SS}, accounting for self-shielding increases the Pop. III SFRD, especially when the Lyman-Werner background starts to build up. At $z \leq 15$ the Pop. III SFRD when the H$_2$ self-shield is considered (cyan dashed line) is $\sim$ 1 order of magnitude higher than the no self-shield scenario (cyan solid). Similar results have been found by \citet{Feathers2024}.

\section{Impact of Pop. III star formation efficiency and IMF}
\label{sec:AppendixB}

\begin{figure}
    %\centering
    \includegraphics[width=\columnwidth]{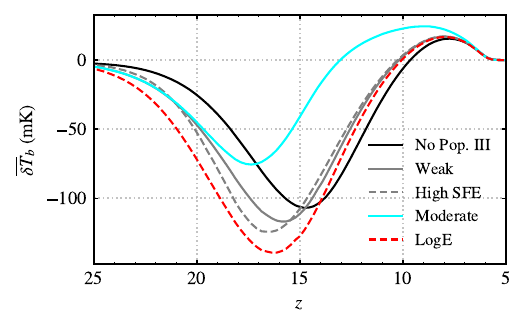}
    \caption{As Fig. \ref{fig:Tb} but showing also the High SFE (grey dashed line) and the LogE (red dashed line) models.}
    \label{fig:Tb_more}
\end{figure}

\begin{figure*}
    \centering
    \includegraphics[width=\textwidth]{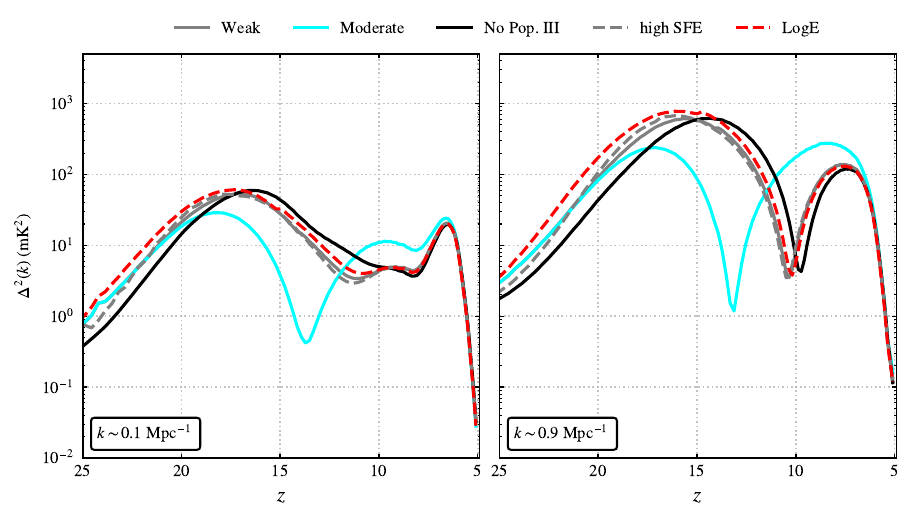}
    \caption{As Fig. \ref{fig:Psvsz} showing also the High SFE (grey dashed line) and the LogE (red dashed line) models.}
    \label{fig:PsvszMore}
\end{figure*}

While the 21cm power spectrum at $z \lesssim 10$ is mostly sensitive to the amount of X-ray photons from Pop. III remnants, changing the star formation efficiency and the shape of the IMF impacts the 21cm signal evolution during the coupling and heating epochs. In Fig. \ref{fig:Tb_more} and \ref{fig:PsvszMore} we show the evolution of the 21cm global signal and power spectrum for the No Pop. III model (black line), Weak (grey line), Moderate (cyan line), high SFE (dashed grey line) and LogE (dashed red line) models. Compared to the Weak Pop. III model, increasing the Pop. III star formation efficiency, increases both the amount of Lyman-$\alpha$ and X-ray photons (they are both linked to SFR), hence the absorption trough shifts towards higher-$z$ ($\Delta z \sim 1$). Considering instead a top-heavy IMF, increases just the amount of Lyman-$\alpha$ photons hence the absorption trough becomes significantly deeper while still being at the same redshift compared to the weak Pop. III model. We note that this result is a consequence of the fact that we are considering the L$_X$/SFR and the shape of the IMF as two independent parameters. If, as shown in \citet{Sartorio2023}, the X-ray emissivity is linked to the shape of the IMF, the LogE model is not realistic as we would need to enhance the X-ray emissivity as well.

We highlight that both the star formation efficiency and the shape of the IMF alone do not significantly affect neither the 21cm global signal nor the power spectrum at $z \lesssim 12$ so a future 21cm detection with SKA-low likely will not help to constrain these parameters if they are not linked to a significantly stronger X-ray emissivity.

%%%%%%%%%%%%%%%%%%%%%%%%%%%%%%%%%%%%%%%%%%%%%%%%%%

% Don't change these lines
\bsp	% typesetting comment
\label{lastpage}
\end{document}